\documentclass[preprint,superscriptaddress,amsmath,amssymb,longbibliography]{revtex4-2}

\usepackage{graphicx}
\usepackage{bm}

\usepackage[hidelinks]{hyperref}
\hypersetup{
    colorlinks,
    linkcolor={blue},
    citecolor={blue},
    urlcolor={blue}
}
\usepackage[dvipsnames]{xcolor}
\renewcommand{\figurename}{Figure}
\definecolor{darkred}{rgb}{0.753,0,0}

\begin{document}

\title{Coherent structures at the origin of time irreversibility in wall turbulence}

\author{Giovanni Iacobello}
    \email{g.iacobello@surrey.ac.uk}
    \affiliation{School of Mechanical Engineering Sciences, University of Surrey, Guildford, GU2 7XH, UK}

\author{Subharthi Chowdhuri}
    \affiliation{Department of Civil and Environmental Engineering, University of California, Irvine, CA 92697, USA}

\author{Luca Ridolfi}
    \affiliation{Department of Environmental, Land and Infrastructure Engineering, Politecnico di Torino, Turin, 10129, Italy}

\author{Lamberto Rondoni}
    \affiliation{Department of Mathematical Sciences, Politecnico di Torino, Turin, 10129, Italy}
    \affiliation{Istituto Nazionale di Fisica Nucleare, INFN, Sezione di Torino, Turin, 10125, Italy}
   
\author{Stefania Scarsoglio}
    \affiliation{Department of Mechanical and Aerospace Engineering, Politecnico di Torino, Turin, 10129, Italy}




\begin{abstract}

    \noindent\textbf{Abstract:} Time irreversibility is a distinctive feature of non-equilibrium phenomena such as turbulent flows, where irreversibility is mainly associated with an energy cascade process. The connection between time irreversibility and coherent motions in wall turbulence, however, has not been investigated yet. An Eulerian, multiscale analysis of time irreversibility in wall-bounded turbulence is proposed in this study, which differs from previous works relying on a Lagrangian approach and mainly focusing on homogeneous turbulence. Outcomes reveal a strong connection between irreversibility levels and coherent structures in both turbulent channel and boundary layer flows. In the near-wall region, irreversibility is directly related to the inner spectral peak originating from small-scale turbulent structures in the buffer layer. Conversely, stronger irreversibility is found in correspondence to the outer spectral peak originating from larger turbulent flow scales far from the wall. Our results represent a first effort to characterize Eulerian TI in wall-bounded turbulent flows, thus paving the way for further developments in wall-turbulence modeling and control accounting for broken temporal symmetry.

\end{abstract}

\maketitle


\section*{Introduction}

    Time irreversibility (TI) is a fundamental property of non-equilibrium systems, which are typically dissipative~\cite{evans1993probability, lebowitz1993boltzmann}. In steady state, TI appears as an asymmetry of the statistical properties of a signal when the time direction is reversed~\cite{cox1991long, Lawrance1991DirectionalityAR, lebowitz1993boltzmann, bertini2001fluctuations, giberti2007temporal}. TI plays an important role in revealing key features of nonlinear dynamical systems, such as long-range nonlinearity and non-Gaussianity, and is inherently associated with entropy production in statistical mechanics and thermodynamics~\cite{cox1991long, porporato2007irreversibility}. Owing to its importance in non-equilibrium systems, TI has been investigated through the lens of time-series analysis in many scientific areas~\cite{costa2005broken, porporato2007irreversibility, zorzetto2018extremes, skinner2021estimating, o2022time}.
    
    Turbulent flows represent a paradigmatic example of dissipative, highly-nonlinear, and far-from-equilibrium systems~\cite{frisch1995turbulence, pope2000turbulent}. A distinctive feature of turbulent flows is the presence of a broad range of scales -- from the largest, inertial, scales to the smallest, dissipative, scales -- across which energy is redistributed~\cite{pope2000turbulent}. The presence of a separation between the large flow scales (where energy is injected) and the smallest scales (where energy is dissipated) implies an average flux of kinetic energy as a cascade process (occurring from large to small length scales in three-dimensional turbulence)~\cite{frisch1995turbulence, pope2000turbulent}. Specifically, far away from solid walls and in a range of scales in-between the large and small scales, the rate at which kinetic energy crosses through the flow scales -- characterizing the cascade process -- can be equated to the turbulent kinetic energy dissipation~\cite{vassilicos2015dissipation}, which has been shown to remain finite even as the viscosity tends to zero (a feature that is referred to as the \textit{dissipative anomaly})~\cite{frisch1995turbulence, jaccod2021constrained}. Accordingly, while the presence of viscosity formally leads to intrinsic irreversibility, high-Reynolds-number turbulence -- owing to its high complexity and the wide large-to-small scale separation -- is mainly driven by inertial forces that make dissipative effects less evident, thereby manifesting significant statistical irreversibility that is typically associated with the energy cascade process~\cite{frisch1995turbulence, xu2014flight, xu2016lagrangian, vela2021entropy}. In dynamical systems theory, the dualism between intrinsic and statistical irreversibility has been widely explored in terms of microscopical and macroscopic irreversibility, such that a dynamical system can be microscopically (intrinsically) reversible and macroscopically (statistically) irreversible~\cite{chibbaro2014reductionism, vela2021entropy}.
    
    \begin{figure*}[t]
        \centering
        \includegraphics[width=0.99\linewidth]{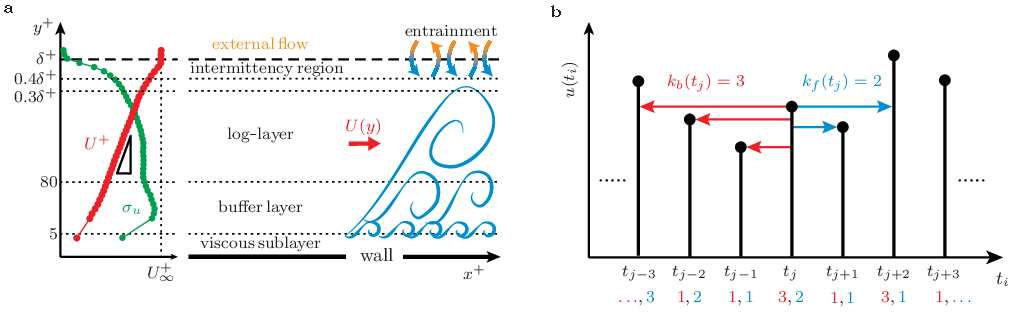}
        \caption{\textbf{Summary sketch of wall turbulence and {horizontal visibility graphs (HVGs)}.} \textbf{a} Schematic of the wall-normal structure of a turbulent boundary layer: (left) vertical profiles of the mean velocity $U^+(y^+)$ (red line, with $U_\infty^+=31.89$) and the root-mean-square velocity $\sigma^+_u(y^+)$ (green line), as a function of the normalized wall-normal coordinate $y^+$ (in a log-scale); (middle) the main vertical layers delimited by conventional $y^+$ limits~\cite{pope2000turbulent}; (right) a qualitative sketch of scale arrangement in the flow. \textbf{b} Sketch of a time series $u(t_i)$ (black vertical lines), and HVG links ({red and blue} colored arrows) for a node $t_j$. Red-blue number pairs indicate backward and forward degree values, $k_{\mathrm{b}}$ and $k_{\mathrm{f}}$ respectively, for each node illustrated as a black-filled circle on top of each vertical line.}
        \label{fig:fig1}
    \end{figure*}

    Several studies have been carried out with the aim to shed light on time irreversibility in turbulent flows and its connection with the energy cascade process, with a particular focus on homogeneous-isotropic turbulence~\cite{xu2014flight, jucha2014time, buaria2015characteristics, bragg2016forward, josserand2017turbulence, cencini2017time} and two-dimensional (2D) turbulence~\cite{rondoni1999fluctuations, gallavotti2004lyapunov, bragg2018irreversibility, porporato2020fluctuation}. In particular, the Lagrangian viewpoint has been recently adopted as a framework to investigate TI, by looking at the asymmetry in the statistics of backward and forward dispersion of tracer particles~\cite{xu2014flight, jucha2014time, buaria2015characteristics, bragg2016forward, xu2016lagrangian, cencini2017time, polanco2018relative, zhang2021time, cheminet2022eulerian}. In contrast, the investigation of TI in wall-bounded turbulent flows has received much less attention~\cite{polanco2018relative, zorzetto2018extremes}, although wall turbulence plays a crucial role in several key engineering and geophysical problems~\cite{jimenez2013near, li2019large}. Wall-bounded turbulence includes an additional level of complexity with respect to homogeneous isotropic turbulence due to the inhomogeneity and anisotropy of the flow along the wall-normal direction. This additional peculiarity of wall turbulence implies that statistical irreversibility in wall turbulence can arise, not only via the energy cascade process but also due to the complex spatio-temporal development of wall-induced coherent motion, which constitutes the backbone of turbulent flows~\cite{jimenez2018coherent}.

    In order to fill the knowledge gap concerning the connection between TI and coherent motion, in this work we investigate statistical irreversibility in wall-bounded turbulence focusing on how different flow scales contribute to TI. Characteristic flow scales, in fact, are associated with coherent structures, and display complex dynamics, not only in terms of energy cascade but also through inter-scale interactions such as modulation phenomena~\cite{baars2015wavelet, iacobello2021large}. {Specifically,} {instead of relying on the Lagrangian viewpoint, we adopt an Eulerian, multiscale framework to investigate TI. In the Lagrangian framework, characteristic scales are identified through the average distance between particles, which is a function of time because particle separation increases (on average) with time~\cite{xu2016lagrangian}. It follows that particle motion senses (Eulerian) turbulent scales of varying sizes at different times due to particle relative dispersion in the flow, thus making it difficult to perform a scale-dependent study of TI. The Lagrangian multiscale analysis is even more challenging in the case of wall turbulence, where the flow dynamics (hence, scales' features) is strongly dependent on the distance from the wall, and Lagrangian dispersion significantly depends on particle wall-normal position~\cite{polanco2018relative}. The Eulerian framework adopted here, instead, allows us to carry out a scale decomposition with ease by exploiting the Fourier transform, as it is typically done in the study of turbulent signals~\cite{smits2011high, jimenez2018coherent, pope2000turbulent}. 
    \\Recent works have provided insights on the relation between Eulerian and Lagrangian TI in turbulent flows, showing a distinct correlation between Eulerian and Lagrangian TI indicators~\cite{cheminet2022eulerian, drivas2019turbulent}. However, they have investigated a turbulent von Kármán flow adopting an instantaneous Eulerian approach~\cite{cheminet2022eulerian}, rather than focusing on wall-bounded turbulence through a multiscale approach. The {key aspect} of our work, therefore, is the analysis of statistical TI at different time scales, and how TI relates to organized motions in the flow. The choice of the flow case represents a {key aspect} as well, since wall turbulence has been much less investigated than homogeneous-isotropic turbulence in terms of statistical TI. Our multiscale analysis, indeed, is particularly {challenging} in wall turbulence due to the complex organization of turbulent structures arising along the wall-normal direction~\cite{jimenez2018coherent}.}

    \begin{figure*}
        \centering
        \includegraphics[width=0.99\textwidth]{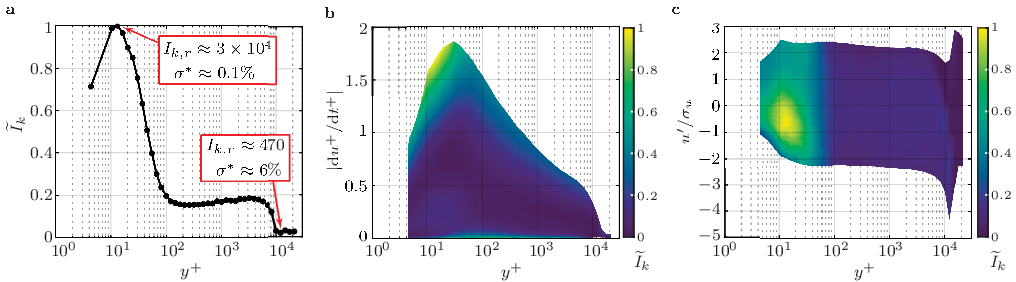}
        \caption{\textbf{Full-signal {time irreversibility (TI)} results for the turbulent boundary layer.} \textbf{a} Wall-normal behavior of $\widetilde{I}_k$. The maximum and minimum values of the irreversibility ratio $I_{k,\mathrm{r}}=(I_k-\mu_{k,\mathrm{r}})/\sigma_{k,\mathrm{r}}$ are also highlighted, where $\mu_{k,\mathrm{r}}$ and $\sigma_{k,\mathrm{r}}$ are computed from an ensemble of 100 randomly-shuffled $u(t_i)$ signals. Conditional analyses of $\widetilde{I}_k$ with respect to $y^+$ and, \textbf{b} the local signal slope $|\mathrm{d}u^+/\mathrm{d}t^+|$ or, \textbf{c} $u^\prime/\sigma_u$. Intervals of $|\mathrm{d}u^+/\mathrm{d}t^+|$ and $u^\prime/\sigma_u$ are binned as $2:2:100$ percentiles.}
        \label{fig:fig2}
    \end{figure*}
    
    {To {perform our investigation}, we take advantage of tools of nonlinear time-series analysis to quantify TI in the Eulerian viewpoint, by exploiting 1-dimensional velocity signals measured at various vertical (i.e., wall-normal) coordinates. Specifically, a parameter-free methodology based on horizontal visibility graphs (HVGs) is employed as the reference approach to capture temporal asymmetry in the signals~\cite{lacasa2012time}. Visibility networks represent a widely-used tool for nonlinear time-series analysis~\cite{zou2019complex}, including turbulent and vortical flows~\cite{iacobello2021review}, which have recently been exploited for TI analysis of both steady and unsteady phenomena~\cite{donges2013testing, suyal2014visibility, schleussner2015indications, lacasa2015time, xie2016time, flanagan2016irreversibility, gonzalez2020arrow}. The choice to adopt an HVG-based metric of TI is dictated by ease of implementation, and by results' robustness as arbitrarily-defined parameters are not required~\cite{lacasa2012time, lacasa2015time}. In fact, one of the main obstacles in quantifying TI is providing robust estimates, which is rarely achievable via traditional time-series symbolization~\cite{lacasa2012time, kennel2004testing}. Visibility graphs, therefore, allow us to directly quantify TI from Fourier-filtered signals for the multiscale analysis, thus connecting our findings with previous results from the literature in terms of velocity spectra.
    \\Nevertheless, like other measures of irreversibility, HVG-based metrics provide partial information on the system~\cite{lacasa2012time}. This is inevitable since a single parameter can not fully represent a complex phenomenon. For this reason, aiming to strengthen our results' reliability, we extend the analysis to two alternative TI metrics: a higher-order (\textit{lag-reversibility}) correlation coefficient~\cite{Lawrance1991DirectionalityAR}, and a measure based on the fluctuation theorem~\cite{porporato2007irreversibility}. The results based on such alternative TI metrics ({see Supplementary Note 3}) corroborate the HVG-based analysis thus advocating its applications.}
    
    {Since coherent structures exhibit different features along the wall-normal direction (e.g., in terms of geometry, dynamics, as well as energetic content~\cite{smits2011high, jimenez2018coherent}), we study the effect of the wall-normal coordinate on TI levels, exploring the whole range of wall-normal distances from the near-wall region (where wall-induced effects are dominant) to the outer flow region (where wall-induced effects are negligible). By so doing, we are able to focus on the relation between TI and flow scales while also accounting for the inhomogeneity and anisotropy of wall-bounded turbulence, which affects the spatio-temporal development of coherent structures. To this aim, experimental and numerical data of the streamwise velocity are exploited from both external (turbulent boundary layer) and internal (turbulent channel) flows at high Reynolds numbers. The streamwise velocity component, $u$, is chosen as the observable for this study since it is well-known to retain fundamental features of wall-bounded turbulence mechanisms; for example, $u$ has about twice as much kinetic energy as the other velocity components, thus making it one of the key variables to consider in wall turbulence analysis~\cite{jimenez2018coherent}.} {Results for both external and internal wall turbulence reveal peculiar scale-dependent patterns, where significant TI levels emerge in correspondence with energetic small and large structures at characteristic wall-normal coordinates. Present outcomes, therefore, suggest a connection between TI and the dynamical processes related to the development of coherent structures in wall turbulence, thus triggering new studies in turbulence research.}

     \begin{figure*}
        \centering
        \includegraphics[width=0.99\linewidth]{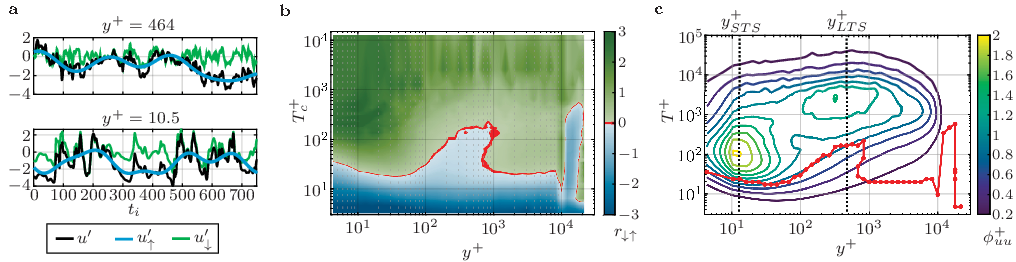}
        \caption{\textbf{Scale-dependent results for the turbulent boundary layer.} \textbf{a} Time series of $u^\prime$ (black) at $y^+=10.5$ and $y^+=464$, and their components $u^\prime_\uparrow$ ({blue}) and $u^\prime_\downarrow$ ({green}) obtained for $T_{\mathrm{c}}^+\approx 170.3$. \textbf{b} Ratio $r_{\downarrow\uparrow}$ as a function of $y^+$ and $T_{\mathrm{c}}^+$. \textbf{c} Energy spectra of $u^\prime$, $\phi_{uu}(T^+,y^+)$, pre-multiplied by frequency $f$ and normalized through $U_\tau^2$ as $\phi_{uu}^+=f\phi_{uu}/U_\tau^2$ (contour level range $0.2-2$, level step $0.2$). $T_{\mathrm{tr}}^+(y^+)$ is shown as a red line, while the two vertical dotted lines refer to $y_{\mathrm{STS}}^+\approx 13$ and $y_{\mathrm{LTS}}^+=3.9\sqrt{R_\tau}$.}
        \label{fig:fig3}
    \end{figure*} 
        
\section*{Results}

    \noindent\textbf{Full-signal TI analysis of turbulent boundary layer.} A laboratory zero-pressure-gradient turbulent boundary layer at friction Reynolds number $R_\tau\approx 14{,}750$ is chosen as the representative (high Reynolds number) test case for inner flows (see a schematic in Fig.~\ref{fig:fig1}a)~\cite{baars2015wavelet}. Streamwise velocity time series, recorded at varying wall-normal distances $y^+$ (the $+$ superscript indicates wall-units normalization by the friction velocity, $u_\tau$, and viscosity, $\nu$), are mapped into directed graphs following the horizontal visibility algorithm~\cite{luque2009horizontal}, as illustrated in Fig.~\ref{fig:fig1}b (see Methods). TI is then quantified as the Kullback-Leibler divergence, $I_k$, of the backward- ($k_{\mathrm{b}}$) and forward-degree ($k_{\mathrm{f}}$) probability distributions, $p(k_{\mathrm{b}})$ and $p(k_{\mathrm{f}})$, where the node degree $k$ quantifies the number of links adjacent to each node (e.g., see Fig.~\ref{fig:fig1}b)~\cite{lacasa2012time, lacasa2015time}. Time reversible signals imply $I_k=0$ (where zero is exactly reached for infinitely-long signals), while growing values $I_k>0$ indicate stronger levels of TI~\cite{lacasa2012time}.

    A full-signal (i.e., not scale-dependent) analysis is carried out first. Figure~\ref{fig:fig2}a shows the behavior of $\widetilde{I}_k=I_k/I_{k,\mathrm{max}}$ (where $I_{k,\mathrm{max}}$ is the maximum $I_k$ value along $y^+$), computed from full-length $u(t_i)$ signals. While a slightly-inclined plateau is observed in the log-layer with a drop in the external region, a $\widetilde{I}_k$ peak is observed in the near-wall region. The emergence of higher TI levels in the buffer layer (which is known to be a very active flow region) is a peculiar result, as this (near-wall) region of wall turbulence is characterized by the development of coherent motion and bursting events~\cite{jimenez2013near, jimenez2018coherent}, as well as modulation mechanisms~\cite{iacobello2021large}, whose complex dynamics can contribute to the generation of statistical TI. 
    
    In order to quantify the degree of reliability of $\widetilde{I}_k$ values, a reliability ratio, $I_{k,\mathrm{r}}$, is also evaluated following Gonz{\'a}lez-Espinoza et al.~\cite{gonzalez2020arrow}: $I_{k,\mathrm{r}}$ corresponds to the Z-score computed with respect to the mean and standard deviation of $I_{k}$ values from an ensemble of randomly-shuffled $u(t_i)$ signals (which are time-reversible). The maximum and minimum values of the reliability ratio (highlighted in Fig.~\ref{fig:fig2}a) are $I_{k,\mathrm{r}}\gg1$, thus allowing one to ascertain with extreme confidence that streamwise velocity signals are irreversible. A parametric analysis on the impact of decreasing the time-series length on the wall-normal behavior of $I_k$ is discussed in {Supplementary Note 1}, while the whole behavior of $I_{k,\mathrm{r}}$ as a function of $y^+$ is reported in {Supplementary Note 2}.

    To shed more light on the origin of TI in full-length $u(t)$ signals, a conditional analysis is performed with respect to: (i) the local slope in the time series, $|\mathrm{d}u^+/\mathrm{d}t^+|$, and (ii) the velocity fluctuations $u^{\prime}/\sigma_u$, where $u^{\prime}(t)=u(t)-U$ ($U$ is the mean velocity), while $\sigma_u$ is the root-mean-square velocity (see Fig.~\ref{fig:fig1}a). Values of $\widetilde{I}_k$ conditioned to the local slope $|\mathrm{d}u^+/\mathrm{d}t^+|$ for different $y^+$ coordinates are reported in Fig.~\ref{fig:fig2}b. The highest levels of TI are detected for intense temporal variations in the buffer layer, which are reminiscent of near-wall bursting events~\cite{jimenez2018coherent}. This link between high TI levels and bursting events is confirmed by the conditional analysis on turbulent fluctuations $u^{\prime}/\sigma_u$ (Fig.~\ref{fig:fig2}c), which points out larger $\widetilde{I}_k$ values residing in the range $-1 < u^{\prime}/\sigma_u < 0$ in the buffer layer. In fact, near-wall bursting events are typically detected as intervals of $u^\prime(t_i)$ starting at $u^{\prime}/\sigma_u=-1$ and ending at $u^{\prime}/\sigma_u<-0.25$~\cite{bogard1986burst, vinuesa2015documentation, tang2016bursting}, in very good agreement with intervals of large $\widetilde{I}_k$ in Fig.~\ref{fig:fig2}c. 
    \\Results from Fig.~\ref{fig:fig2} are also in qualitative accordance with previous studies where irreversibility was related to (i) the velocity-gradient tensor perceived by dispersing Lagrangian particles~\cite{jucha2014time}, and (ii) an asymmetry between growth and decay of signal fluctuations (i.e., where local signal slopes are large) in non-equilibrium systems~\cite{luchinsky1997irreversibility, bertini2001fluctuations, giberti2007temporal}.

    \begin{figure*}
        \centering
        \includegraphics[width=0.99\linewidth]{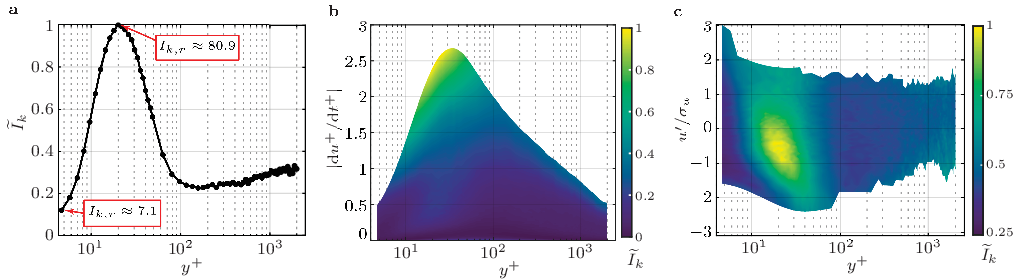}
        \caption{\textbf{Full-signal {time irreversibility (TI)} results for the turbulent channel flow.} \textbf{a} Wall-normal behavior of $\widetilde{I}_k$. The maximum and minimum values of the irreversibility ratio $I_{k,\mathrm{r}}$ are also highlighted. Conditional analysis of $\widetilde{I}_k$ with respect to $y^+$ and \textbf{b} the local signal slope $|\mathrm{d}u^+/\mathrm{d}t^+|= U^+|\mathrm{d}u^+/\mathrm{d}x^+|$, and \textbf{c} $u^\prime/\sigma_u$. Intervals of $|\mathrm{d}u^+/\mathrm{d}t^+|$ and $u^\prime/\sigma_u$ are binned as $2:2:100$ percentiles.}
        \label{fig:fig4}
    \end{figure*}    

    \vspace{10pt}
    \noindent\textbf{Multiscale analysis of TI.} A scale-dependent investigation is carried out here to highlight how different turbulent scales contribute to TI of the streamwise velocity. Accordingly, $u(t_i)$ signals are Fourier-filtered at various cut-off periods $T_{\mathrm{c}}^+=1/f_{\mathrm{c}}^+$ such that $u(t_i)= u_\uparrow(t_i)+ u_\downarrow(t_i)$, where $u_\uparrow$ and $u_\downarrow$ indicate the low-pass and high-pass filtered velocity components of $u$, respectively (see Fig.~\ref{fig:fig3}a). The HVG-algorithm is then applied to the $u_\uparrow$ and $u_\downarrow$ velocity components, thus obtaining the corresponding TI measures $I_{k,\uparrow}$ and $I_{k,\downarrow}$, respectively. Figure~\ref{fig:fig3}b shows the ratio 
        
    \begin{equation}
        r_{\downarrow\uparrow} = \log_{10}{\frac{I_{k,\downarrow}}{I_{k,\uparrow}}},    
    \end{equation}
        
    \noindent as a function of $y^+$ and $T_{\mathrm{c}}^+$: blue-shaded areas ($r_{\downarrow\uparrow}<0, I_{k,\downarrow}<I_{k,\uparrow}$) indicate a greater contribution to TI coming from temporal scales of $u(t)$ with larger periods than $T_{\mathrm{c}}^+$; green-shaded areas ($r_{\downarrow\uparrow}>0, I_{k,\downarrow}>I_{k,\uparrow}$) instead indicate a greater contribution to TI coming from temporal scales of $u(t)$ with smaller periods than $T_{\mathrm{c}}^+$. It should also be pointed out that, in the limiting cases of $T_{\mathrm{c}}^+\gg 0$ (bounded by Nyquist frequency) and $T_{\mathrm{c}}^+\rightarrow 0$, all temporal scales will contribute either to $I_{k,\downarrow}$ (hence $r_{\downarrow\uparrow}\gg 0$) or $I_{k,\uparrow}$ (hence $r_{\downarrow\uparrow}\ll 0$).

    The transitional line $T_{\mathrm{tr}}^+(y^+)$ for which $r_{\downarrow\uparrow}\approx 0$ (i.e., $I_{k,\downarrow}\approx I_{k,\uparrow}$, see red contour in Fig.~\ref{fig:fig3}b) discriminates between the region where the cut-off filter is $T_{\mathrm{c}}^+<T_{\mathrm{tr}}^+$ (such that $r_{\downarrow\uparrow}<0$), and the region where the cut-off filter is $T_{\mathrm{c}}^+>T_{\mathrm{tr}}^+$ (such that $r_{\downarrow\uparrow}>0$). The $T_{\mathrm{tr}}^+(y^+)$ boundary displays a peculiar behavior, made up of two bumps emerging far from the wall. The first bump arises at $y^+\approx 450$ in the log-layer, while an additional bump is present at $y^+\approx 14{,}500$ in the intermittency region (see the sketch in Fig.~\ref{fig:fig1}a), which is the flow layer where the entrainment process occurs owing to the proximity with the external (non-turbulent) flow~\cite{pope2000turbulent}. The occurrence of bumps in the $T_{\mathrm{tr}}^+(y^+)$ behavior suggests the presence of relevant phenomena in the flow, leading to higher TI levels at larger temporal scales (since higher cut-off periods $T_{\mathrm{c}}^+=T_{\mathrm{tr}}^+$ are needed to have the balance $I_{k,\uparrow}= I_{k,\downarrow}$, i.e., $r_{\downarrow\uparrow}=0$).

    \vspace{10pt}
    \noindent\textbf{Relating TI levels and turbulent flow scales.} With the aim to connect the $r_{\downarrow\uparrow}$ behavior (specifically, the presence of bumps in $T_{\mathrm{tr}}^+(y^+)$) with the underlying structure of the turbulent boundary layer, we show in Fig.~\ref{fig:fig3}c the pre-multiplied energy spectrum of $u$, $\phi_{uu}^+$. Two peaks are distinguishable in Fig.~\ref{fig:fig3}c around $y_{\mathrm{STS}}^+\approx 13$ (buffer layer) and $y_{\mathrm{LTS}}^+\approx 470$ (log-layer): they are associated with the development of organized coherent flow structures commonly referred to as \textit{small turbulent scales} (STS) and \textit{large turbulent scales} (LTS), respectively, the latter emerging at high Reynolds numbers~\cite{smits2011high, baars2015wavelet, iacobello2021large}.

    The $T_{\mathrm{tr}}^+(y^+)$ line (i.e., where $I_{k,\uparrow}=I_{k,\downarrow}$) is also displayed (in red) in Fig.~\ref{fig:fig3}c, showing that both STS and LTS spectral peaks stand above $T_{\mathrm{tr}}^+(y^+)$. Hence, for a cut-off filter $T_{\mathrm{c}}^+$ such that $T_{\mathrm{c}}^+\equiv T_{\mathrm{tr}}^+$, the contribution to TI coming from STS and LTS is -- in both cases -- enclosed in $I_{k,\uparrow}$ since, by definition, $I_{k,\uparrow}$ accounts for the contribution to TI coming from flow scales larger than $T_{\mathrm{c}}^+= T_{\mathrm{tr}}^+$. By increasing the value of the cut-off period, i.e. $T_{\mathrm{c}}^+>T_{\mathrm{tr}}^+$, a growing contribution to TI coming from flow scales associated with STS and LTS is also enclosed in $I_{k,\downarrow}$, leading to $I_{k,\downarrow}>I_{k,\uparrow}$ (i.e., $r_{\downarrow\uparrow}>0$). Therefore, the presence of energetic coherent structures in the flow (STS and LTS) significantly affect the levels of TI of $u(t)$ at varying wall-normal locations, as captured by the relative intensity of $I_{k,\downarrow}$ and $I_{k,\uparrow}$.
    \\Furthermore, we note that a larger extent of dark-green areas (i.e., $r_{\downarrow\uparrow}\geq 1$ or $I_{k,\downarrow}/I_{k,\uparrow}\geq 10$) emerges in Fig.~\ref{fig:fig3}b closer to the $T_{\mathrm{tr}}^+(y^+)$ boundary at $y_{\mathrm{STS}}^+$ than at $y_{\mathrm{LTS}}^+$. The different contribution to TI at various $y^+$ can hence be related to the different intensities of energetic peaks, as the inner peak at $y_{\mathrm{STS}}^+$ is stronger than the outer peak at $y_{\mathrm{LTS}}^+$. This resonates with the previous results of Fig.~\ref{fig:fig2} where larger levels of TI were observed in the buffer layer from a full-signal perspective. This outcome corroborates the key role of the buffer layer as a region where complex dynamical processes are at play and which lead to greater TI.

    Concerning the intermittency region (see Fig.~\ref{fig:fig1}a), previous works~\cite{chauhan2014turbulent} found that a characteristic (large) length scale of entrainment is $\Lambda_x^+\approx 1.7\,\delta^+$. The $T_{\mathrm{tr}}^+$ bump in this region occurs at $T^+\approx 600$ and corresponds to a length scale (using Taylor hypothesis) of $600\,U_\infty^+\approx 1.3\,\delta^+$, which is -- similarly to STS and LTS -- close to the characteristic flow scale $1.7\,\delta^+$ although slightly smaller. The present result reveals that, despite very low TI levels are detected from a full-signal analysis in the intermittency region (see the right tail of Fig.~\ref{fig:fig2}a), a significant link between TI and the entrainment process can be established.

    \vspace{10pt}
    \noindent\textbf{TI analysis of turbulent channel flow.} The close relation between TI and the arrangement of coherent flow motion in the turbulent boundary layer is further validated here by using a second test case, consisting of a turbulent channel flow at $R_\tau=2{,}003$~\cite{hoyas2006scaling}. Streamwise velocity signals $u(x)$ are extracted from a direct numerical simulation (DNS) extensively validated by previous studies~\cite{hoyas2006scaling, jimenez2008turbulent}. The common assumption of Taylor's hypothesis is adopted to investigate spatial signals $u(x)$ as equivalent time series $u(t)$ via HVGs~\cite{smits2011high, iacobello2021large}. The use of spatial series here is related to two main arguments. From one side, the output of numerical simulations is typically a sequence of (time) snapshots of the spatial data, thereby the extraction of long time series from DNSs is computationally challenging for large Reynolds number flows. On the other side, the use of spatial series in our work allows us to provide some insights about the time-space issue -- i.e., to what extent the temporal turbulence dynamics relates to the corresponding spatial dynamics -- in wall turbulence~\cite{squire2017applicability}, specifically in terms of applicability of Taylor’s hypothesis in the context of time irreversibility.
    
    Figure~\ref{fig:fig4} reports the same quantities as shown in Fig.~\ref{fig:fig2} for the turbulent boundary layer, highlighting that results for the turbulent channel flow are in agreement with those for the turbulent boundary layer. In particular, in Fig.~\ref{fig:fig4}a, we observe a peak of $\widetilde{I}_k$ in the buffer layer (as for the turbulent boundary layer) with values of the confidence ratio $I_{k,\mathrm{r}}>1$ (highlighted by red boxes). However, due to the lack of an intermittency region for the channel flow, we do not observe any drop of $\widetilde{I}_k$ for $y^+\rightarrow y^+_{\mathrm{max}}=2{,}003$ in Fig.~\ref{fig:fig4}a. Moreover, results from the conditional analysis in the channel flow (Fig.~\ref{fig:fig4}b-c) match those for the boundary layer flow (Fig.~\ref{fig:fig2}b-c), with a peak of $\widetilde{I}_k$ for strong local slopes $|\mathrm{d}u^+/\mathrm{d}t^+|$ (\figurename~\ref{fig:fig4}b) and for $u^\prime/\sigma_u\approx -0.5$ (\figurename~\ref{fig:fig4}c) in the buffer layer.

    \begin{figure}
        \centering
        \includegraphics[width=0.5\linewidth]{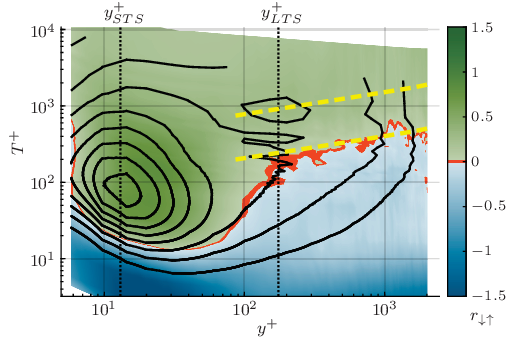}
        \caption{\textbf{Scale-dependent results for the turbulent channel flow.} Colored areas corresponds to the ratio $r_{\downarrow\uparrow}$, while the black contours refers to the pre-multiplied energy spectra $\phi_{uu}^+$ (contour level range $0.25-2$, level step $0.25$). The two yellow dashed lines refer to the $\lambda_x^+\sim{y^+}^{3/7}$ scaling~\cite{monty2009comparison}, in terms of time scales $T^+=\lambda_x^+/U^+$.}
        \label{fig:fig5}
    \end{figure}

    Figure~\ref{fig:fig5} shows the energy spectra (black contours) together with the ratio $r_{\downarrow\uparrow}$ (colored area), where $I_{k,\downarrow}$ and $I_{k,\uparrow}$ are evaluated for different cut-off periods $T_{\mathrm{c}}^+=\lambda_x^+/U^+$ (where $\lambda_x$ is the wavelength along $x$). Figure~\ref{fig:fig5} is in excellent agreement with outcomes from Fig.~\ref{fig:fig3}b-c, highlighting larger green-shaded areas in correspondence of the spectral peak in the buffer layer at $y^+_{\mathrm{STS}}$. The notable difference is the absence of bumps at larger $y^+$, due to two main reasons. First, the intermittency region is absent in a channel flow, hence the rightmost bump of Fig.~\ref{fig:fig3}b does not appear in Fig.~\ref{fig:fig5}. Second, the bump in correspondence of $y^+_{\mathrm{LTS}}$ does not emerge because very large-scale flow motions extend much further towards higher $y^+$ in channel flows than in boundary layers, although their spectral energy similarly decreases for $y^+\rightarrow y^+_{\mathrm{max}}$~\cite{monty2009comparison}. This statement is supported by scaling arguments of $r_{\downarrow\uparrow}$ in the log-layer. In fact, the transitional (red) line $T_{\mathrm{tr}}^+$ in Fig.~\ref{fig:fig5} follows the same ${y^+}^{3/7}$ scaling law (see dashed yellow lines in Fig.~\ref{fig:fig5}) found for the size of very large-scale motion in internal wall-bounded turbulence~\cite{monty2009comparison}. This outcome further corroborates the strong connection between time irreversibility and the spatio-temporal development of coherent structures in the flow.

\section*{Discussion}

    The present study provides a novel perspective into the statistical irreversibility of high-Reynolds-number turbulent flows, that complements previous works looking at broken temporal symmetry in the statistics of Lagrangian tracers. In particular, this work represents a first effort to characterize TI from an Eulerian viewpoint in wall-bounded turbulent flows, which have been much less investigated than other flow configurations from the TI point of view. Exploiting tools of nonlinear time-series analysis, we are able to quantify TI from Eulerian data by explicitly highlighting the contributions to TI coming from various flow scales, as well as the effect of the wall-normal (spatial) coordinate, revealing non-trivial TI patterns.

    Our findings -- relying on the visibility graph-based approach and corroborated through additional methodologies, as reported in {Supplementary Note 3} -- point out that broken temporal symmetry in the streamwise velocity is significantly linked to the underlying (space-dependent) organized flow structure of wall turbulence. In fact, we show that the scale-dependent TI levels follow non-monotonic trends at all wall-normal coordinates, highlighting a non-trivial contribution to TI from smaller and larger scales. We find that TI in the proximity of the wall is dominated by small turbulent scales associated with the inner spectral-energy peak, thus suggesting a connection between TI and the dynamical process related to the near-wall (regeneration) cycle~\cite{jimenez2018coherent}. This claim is partially supported by the large TI levels detected in this work in correspondence with burst events in the buffer layer (see Fig.s~\ref{fig:fig2}b-c and \ref{fig:fig4}b-c). Moving away from the wall, the largest contribution to TI shifts from small to large turbulent scales, which appear at a high Reynolds number and make a significant contribution to the turbulent kinetic energy and Reynolds stress production~\cite{smits2011high, jimenez2018coherent}. Moreover, significant levels of time irreversibility in the intermittency region of turbulent boundary layers are associated with the characteristic scale of the entrainment process. 
    \\It is important to underline here that results from full velocity signals (Fig.~\ref{fig:fig2}a and Fig.~\ref{fig:fig4}a) do not highlight the specific contribution to TI from different flow scales, but they point out that TI levels are larger in the buffer layer. The scale-dependent analysis (Fig.~\ref{fig:fig3} and Fig.~\ref{fig:fig5}), instead, is able to reveal the origin of TI levels in terms of flow scales at various $y^+$, e.g., highlighting the role of large turbulent scales in the log-layer that is not clearly detectable through the full-signal analysis.
   
    In this picture, TI patterns in the near-wall region are in agreement between the turbulent boundary layer (external flow) and the turbulent channel (internal flow), while dissimilarities emerge far from the wall due to intrinsic differences in the flow features of external and internal wall turbulence (e.g., the absence of an intermittency region in internal flows). Overall, the outcomes for both turbulent channel and turbulent boundary layer flows indicate that TI patterns are not limited to a particular flow case, but are a distinctive feature of wall-bounded turbulent flows at high Reynolds number.
    
    It is worth noting here that the full-signal time irreversibility behavior shown in Fig.s~\ref{fig:fig2}a and \ref{fig:fig4}a displays a wall-normal trend similar to the skewness of the spatial derivative $\partial u/\partial x$ reported in previous works~\cite{vreman2014statistics, djenidi2017skewness}, with important implications. Such a similarity suggests that $\partial u/\partial x$ can be reinterpreted as a surrogate measure of TI, although $\partial u/\partial x$ is computed at a fixed time. However, the TI reinterpretation of $\partial u/\partial x$ is conditional to the applicability of Taylor's hypothesis, which allows the conversion of spatial signals $u(x)$ into equivalent time series $u(t)$, and \textit{vice versa}. This point is crucial because an unconsidered use of spatial asymmetry metrics to quantify TI may lead to inconsistent outcomes, especially for non-canonical flow setups~\cite{zorzetto2018extremes}. For the flow cases considered in this study, the application of classical Taylor's hypothesis -- namely, when the convection velocity equals the mean velocity -- led to consistent outcomes from time- and spatial-series, as discussed for the turbulent channel flow case, but caution is still needed. In this sense, refined formulations of Taylor's hypothesis -- e.g., accounting for modulation mechanisms in wall turbulence~\cite{iacobello2021large} -- could help to shed more light on the duality between temporal and spatial asymmetries and will be explored in future works. 
    \\Another thought-provoking similarity appears between the vertical behavior of $I_k$ (Fig.s~\ref{fig:fig2}a and \ref{fig:fig4}a) and the vertical behavior of the Corrsin integral parameter in wall turbulence (as reported, e.g., {by~\citet{jimenez2018coherent}}); the latter is the ratio between the characteristic timescale of energetic turbulence eddies and the characteristic timescale due to the mean shear~\cite{jimenez2018coherent}. The qualitative similarity between these two quantities could suggest a key role played by mean shear in the mechanisms leading to higher TI levels, thus turning the spotlight on the consideration of other phenomena (than the energy cascade) affecting statistical TI and stimulating further research on TI in shear flows.
    
    Furthermore, it should be pointed out that different metrics need to be examined and compared in order to provide reliable TI outcomes (as done in this study; see {Supplementary Note 3}). Since TI can be quantified using different tools~\cite{Lawrance1991DirectionalityAR, lacasa2012time}, the choice of the methodologies can be problem dependent but should rest upon the method interpretability and robustness (e.g., avoiding subjective binning or symbolization procedures whenever possible). In light of this, visibility graphs represent a versatile and robust approach that deserves further consideration for the investigation of turbulence signals~\cite{iacobello2021review}, even beyond the streamwise velocity component considered here. In this regard, the comparison of the results from different observables (e.g., different velocity components) deserves further investigation with the goal of providing a more comprehensive picture of TI in wall turbulence, a topic left to be explored in future works.

    In conclusion, the key finding of the present Eulerian analysis is the emergence of distinctive, scale-dependent, patterns of TI in wall turbulence, originating in correspondence with the development of small and large energetic structures at various, characteristic, wall-normal distances. To the best of our knowledge, present results are new in the context of turbulence research and represent a conceptual advancement in the characterization of statistical TI in wall turbulence, specifically by providing further insights on the relation between TI and the multi-scale arrangement of turbulence at various wall-normal distances. This work leaves some issues open -- e.g., what is the role of mean shear and what is the impact of Taylor's hypothesis in TI quantification -- that can trigger new studies in turbulence research.

\section*{Methods}    
    
    \noindent\textbf{Horizontal visibility graphs and TI quantification.} Given a time series $u(t_i)$ evaluated at discrete times $t_i$, a horizontal visibility graph representation of $u(t_i)$ is obtained, (i) assigning each time $t_i$ to a network node, and (ii) linking two nodes $t_i$ and $t_j$ when the conditions
        \begin{equation}
            u(t_i)>u(t_l) \wedge u(t_j)>u(t_l),\label{eq:hvg_def}
        \end{equation}
        
        \noindent are satisfied for all $t_l$ such that $t_i<t_l<t_j$~\cite{luque2009horizontal}. Consecutive data points, $u(t_i)$ and $u(t_{i+1})$, are always linked by construction (i.e., the index $l$ can be null). The resulting graph is stored as a binary adjacency matrix $A_{ij}$, whose entries are $A_{ij}=1$ if and only if nodes $i$ (corresponding to datum $u(t_i)$) and $j$ (corresponding to datum $u(t_j)$) are linked with each other as per Eq.~(\ref{eq:hvg_def}). Horizontal visibility networks are typically constructed as undirected graphs, namely the link direction is not taken into account, and $A_{ij}=A_{ji}$ for any node pair $(i,j)$. In this work, however, we consider directed graphs, namely explicitly differentiating between forward-in-time links $A_{ij}=1$ (i.e., when $t_j>t_i$) and backward-in-time links $A_{ji}=1$ (i.e., when $t_j<t_i$). Figure~\ref{fig:fig1}b shows an exemplifying discrete signal, where HVG links for node $t_j$ are represented by colored arrows. 
        
        The amount of links of each node is referred to as degree centrality $k(t_i)$. Here, we distinguish between forward-degree $k_{\mathrm{f}}$ and backward-degree $k_{\mathrm{b}}$ (commonly referred to as out- and in-degree~\cite{zou2019complex}) as the number of links pointing towards increasing and decreasing time, respectively (Fig.~\ref{fig:fig1}b). \citet{lacasa2012time} proposed to quantify TI in a time series as the Kullback-Leibler divergence of the backward- and forward-degree distributions:
        
        \begin{equation}
            I_k=\sum_{k_{\mathrm{b}},k_{\mathrm{f}}}{p(k_{\mathrm{b}})\log{\frac{p(k_{\mathrm{b}})}{p(k_{\mathrm{f}})}}}, \label{eq:Ik_def}
        \end{equation}
        
        \noindent where $p(k_{\mathrm{b}})$ and $p(k_{\mathrm{f}})$ are the marginal probability distributions of $k_{\mathrm{b}}$ and $k_{\mathrm{f}}$, respectively. The degree of reliability of $I_k$ can be assessed by evaluating the \textit{irreversibility ratio} 
        \begin{equation}
            I_{k,\mathrm{r}}=\frac{I_k-\mu_{k,\mathrm{r}}}{\sigma_{k,\mathrm{r}}},    
        \end{equation}
        \noindent where $\mu_{k,\mathrm{r}}$ and $\sigma_{k,\mathrm{r}}$ are the mean and standard deviation, respectively, of $I_k$ values calculated from an ensemble of signals obtained through a random (null) model~\cite{gonzalez2020arrow}. As random signals by definition privilege no direction, $I_{k,\mathrm{r}}\gg 1$ signifies that a time series is TI with extreme confidence~\cite{gonzalez2020arrow}.

        It should be noted that, in general, the Kullback-Leibler divergence is not a symmetric measure, namely $\sum{ p(k_{\mathrm{b}})\log\left[p(k_{\mathrm{b}})/p(k_{\mathrm{f}})\right]}\neq \sum{p(k_{\mathrm{f}})\log\left[p(k_{\mathrm{f}})/p(k_{\mathrm{b}})\right]}$. However, switching the position of $p(k_{\mathrm{b}})$ and $p(k_{\mathrm{f}})$ in Eq.~(\ref{eq:Ik_def}) does not significantly change the outcomes and hence our conclusions (as also reported in previous studies~\cite{lacasa2012time,lacasa2015time}). Moreover, $I_k$ is not an additive measure due to the non-linearity of the Kullback-Leibler divergence, thereby the sum $I_{k,\uparrow}+I_{k,\downarrow}$ (obtained through high-pass and low-pass filtering of the signal) is in general not equal to $I_k$ (obtained from the full signal).
        
        \vspace{10pt}
        \noindent\textbf{Turbulent boundary layer experiments.} The boundary layer was experimentally obtained in the wind-tunnel facility of the University of Melbourne~\cite{MARUSIC2020dataset}. The friction Reynolds number is $R_\tau= \delta U_\tau/\nu \approx 14{,}750$, where $\delta= 0.361$\,m and $U_\tau = 0.626\,\mathrm{m\,s^{-1}}$ are the boundary layer thickness and the friction velocity, respectively, while $\nu= 1.532\times 10^{-5}\, \mathrm{m^2s^{-1}}$ is the kinematic viscosity of air. The value of the Reynolds number of the experiment is large enough to ensure a wide range of temporal scales, thereby allowing very-large-scale motions to develop~\cite{baars2015wavelet}. Further details of the experiments can be found in Baars et al.~\cite{baars2015wavelet}.
        
        The time series of $u(t_i)=u^\prime(t_i)+U$ -- where $U(y)$ is the local (time-averaged) mean velocity, while $u^\prime$ are turbulent fluctuations (Fig.~\ref{fig:fig1}a) -- were recorded at $41$ wall-normal coordinates $y$, while fixing the streamwise and spanwise coordinates, $x$ and $z$, respectively. Wall-units normalization is indicated through the $+$ superscript such that $u^+=u/U_\tau$ and $y^+=y U_\tau/\nu$. Time series were collected for three acquisition cycles of $120$\,s at sampling frequency $f_s=20$\,kHz, thereby results are intended to be averaged over the three acquisition cycles.

        \vspace{10pt}
        \noindent\textbf{Turbulent channel numerical simulation.} Data from a direct numerical simulation of a turbulent channel flow at $R_\tau= 2{,}003$ are used as a representative case of internal flow. The resulting dataset is available online on the webpage of the Fluid Dynamics Group of Universidad Politecnica de Madrid ({see Data availability statement})~\cite{hoyas2006scaling}. The simulation is run on a smooth-wall channel setup with periodic boundary conditions in the streamwise ($x$) and spanwise ($z$) directions. The domain size is $8\pi \delta \times 2\delta \times 3\pi \delta$ in the streamwise, vertical, and spanwise directions respectively, where $\delta$ is the half-channel height. The numerical grid consists of $6{,}144$ and $4{,}608$ uniformly-spaced grid points in the streamwise and spanwise direction, respectively, while a non-uniform grid with $633$ points is used in the wall-normal direction. Further details can be found in Hoyas and Jim{\'e}nez~\cite{hoyas2006scaling}.

       Spatial series $u(x)$ are used in this work to quantify TI, where Taylor's hypothesis $\Delta x=-U(y)\Delta t$ is used to transform the spatial signal into time series~\cite{pope2000turbulent, jimenez2018coherent}. In this regard, the spatial resolution $\Delta x^+=8.2$ is equivalent to the sampling time step of the turbulent boundary layer. It should be noted here that, the application of Taylor's hypothesis does not affect the computation of $I_k$ from spatial signals $U(x)$ in the channel flow because the horizontal visibility algorithm is insensitive to constant re-scaling of the horizontal axis (more generally, it is insensitive to affine transformations)~\cite{luque2009horizontal}. One time snapshot from the database is used, thereby ensemble-averages of the results are only performed along the spanwise (homogeneous) direction $z$, over $288$ uniformly-spaced locations. Further increases in the averaging sample size do not substantially change the discussed outcomes.

\section*{Data Availability}
All the data analyzed in this paper are openly available at locations referenced herein. {Turbulent boundary layer: \url{https://doi.org/10.26188/5e919e62e0dac}; turbulent channel flow: \url{https://torroja.dmt.upm.es/turbdata/}.}

\section*{Code Availability}
{Horizontal visibility graphs (HVG) are built using an in-house MATLAB code, which is available online at the following repository: MATLAB Central File Exchange, code reference "72889-fast-horizontal-visibility-graph-hvg-for-matlab", Retrieved April 11, 2023.}

%
\def\bibsection{\section*{References}}
\bibliography{biblio_rev}


\section*{Acknowledgments}
S.C. acknowledges the funding support from the Department of Civil and Environmental Engineering, University of California, Irvine. L.Ro. acknowledges the support of Italian National Group of Mathematical Physics (GNFM) of INDAM, and of Ministero dell’Istruzione e dell’Università e della Ricerca (MIUR), Italy; Grant No. E11G18000350001 ‘‘Dipartimenti di Eccellenza 2018–2022’’.

This version of the article has been accepted for publication, after peer review but is not the Version of Record and does not reflect post-acceptance improvements, or any corrections. The Version of Record is available online at: https://doi.org/10.1038/s42005-023-01215-y.

\section*{Author contributions}

G.I. conceived the study. G.I. and S.C. collected and processed the data. G.I. wrote the initial manuscript and produced the figures with input from S.C., Lu.R. and S.S. Critical feedback on the methodology was provided by La.R. All authors contributed to the design of the paper and its revision, as well as to the interpretation of the results.

\section*{Competing interests}
The authors declare no competing interests.

\end{document}